# Thermo-acoustic wave propagation and reflection near the liquid-gas critical point

P. Zhang[*], B. Shen

Institute of Refrigeration and Cryogenics, Shanghai Jiao Tong University, Shanghai, 200240 China

Abstract:

We study the thermo-acoustic wave propagation and reflection near the liquid-gas critical point. Specifically, we perform a numerical investigation of the acoustic responses in a near-critical fluid to thermal perturbations based on the same setup of a recent ultrasensitive interferometry measurement in $CO_2$ [Y. Miura et al. Phys. Rev. E **74**, 010101(R) (2006)]. The numerical results agree well with the experimental data. New features regarding the reflection pattern of thermo-acoustic waves near the critical point under pulse perturbations are revealed by the proper inclusion of the critically diverging bulk viscosity.

Singularities such as the vanishing thermal diffusivity and diverging compressibility near the critical point (CP) are believed to be responsible for a novel thermo-mechanical phenomenon—the so-called piston effect (PE), which is widely touted as the fourth mode of heat transfer in addition to thermal conduction, convection, and radiation. Unwittingly discovered in a microgravity experiment in the 1980s [1], the PE was later theorized by several independent groups [2]. Claimed to significantly enhance a near-critical fluid's ability to transport energy—the *critical speeding up*, the PE has since drawn considerable attention among critical dynamicists and spawned numerous studies on theoretical, numerical, and experimental fronts [3-6]. One of the basic features is that the temperature relaxation can be realized within a much shorter time $t_{PE}$ (the PE time scale, expressed as $t_{PE}=t_d/(\gamma-1)^2$ [2]), than the characteristic time of pure heat diffusion $t_d$ ($=L^2/D$, with $L$ the characteristic length of the fluid and $D$ the thermal diffusivity) because the ratio between the isobaric and isochoric specific heats ($\gamma=C_P/C_V$) diverges near the CP. The above definition of $t_{PE}$ implies a nonphysical notion of instantaneous relaxation if one moves infinitely close to the CP. Zappoli et al. [7] proposed an *acoustic saturation* to address this anomaly, which confines the thermalization to the order of the acoustic time ($t_{ac}=L/c$, where $c$ denotes the sound velocity). Such an effect, however, was found to be masked by the diverging bulk viscosity [8] and fluid effusivity [9]. A “viscous regime” appears on approaching the CP instead where boundary attenuation is suppressed and bulk damping dominates [9, 10]. Notwithstanding these new findings, to the authors’ knowledge, little is known on the acoustic time scale including the impact of the diverging bulk viscosity on the fast acoustic process, which was identified to cause the fluid dynamical wave behavior of the PE [11].

[*] Email: zhangp@sjtu.edu.cn

In this Rapid Communication, we develop a real-fluid model taking into account all the relevant fluid characteristics near the CP, including the diverging bulk viscosity. To verify its validity, a numerical simulation that aims at reproduction of a high-speed observation of thermo-acoustic wave emission and propagation in near-critical $CO_2$ conducted by Miura et al. [12] is carried out, which captures the wave features on the time scale of microseconds. We adopt the physical configuration proposed by Carlès [13], which showed great promise in interpreting Miura et al.’s results. Specifically, a one-dimensional (1-D) computational domain of length $L$ (i.e., the height of the test cell) is employed. We impose non-slip and isothermal temperature conditions on both ends. The membrane heater immersed in the fluid is modeled by an external energy source with a finite width of 3 $\mu$m placed $d$=0.5 mm to the left boundary. The heater is on for a time duration $t_h$, during which a prescribed amount of heat is released into the fluid in accordance with the experimental parameters. The heating can thus be defined as, $r(t)=\Phi_{max}[1-H(t-t_h)]$, where $H$ is the Heavyside function and $\Phi_{max}$ denotes the heat input per unit time per unit volume. Despite the apparent terrestrial conditions in which the experiment was conducted, gravity is excluded from consideration for the purpose of simplicity.

Under the continuum approximation, the momentum and energy transport can be readily depicted by

$$\frac{\partial \rho}{\partial t} + \nabla \cdot (\rho \mathbf{v}) = 0, \tag{1}$$

$$\frac{\partial (\rho \mathbf{v})}{\partial t} + \nabla \cdot \left( P\mathbf{G} + \rho \mathbf{v} \otimes \mathbf{v} - 2\eta \boldsymbol{\sigma} - \zeta \theta \mathbf{G} \right) = 0, \tag{2}$$

$$\frac{\partial (\rho s)}{\partial t} + \nabla \cdot \left( \rho s \mathbf{v} - \lambda \nabla \ln T \right) = \lambda \left( \nabla \ln T \right)^2 + \frac{2\eta \boldsymbol{\sigma} : \boldsymbol{\sigma} + \zeta \theta^2}{T} + \frac{r}{T}, \tag{3}$$

where the operator $\otimes$ means a tensor product, the colon a double contraction of shear tensor $\boldsymbol{\sigma}$ with itself, and **G** the metric tensor. The above equations are, in tensorial notation, the conservation laws of mass, momentum, and the evolution equation of entropy, respectively. The contributions of thermal conduction and viscous dissipation are defined in terms of the thermal conductivity $\lambda$, and both of the coefficients of bulk and shear viscosity denoted by $\zeta$ and $\eta$. Based on the Newtonian fluid assumption, the viscous stress in the fluid is linearly related to the expansion rate $\theta$ and shear tensor $\boldsymbol{\sigma}$. The term $r/T$ on the right-hand side of Eq. (3) represents the heater. By substituting the first law of thermodynamics and the continuity equation [Eq. (1)] into Eq. (3), the energy equation becomes

$$\rho C_v \frac{dT}{dt} + \frac{\rho\left(C_p - C_v\right)}{\alpha_P} \nabla \cdot \mathbf{v} = \lambda \nabla^2 T + 2\eta \boldsymbol{\sigma} : \boldsymbol{\sigma} + \zeta \theta^2 + r. \qquad (4)$$

where $d/dt \equiv \partial/\partial t + \mathbf{v} \cdot \nabla$ denotes convective time derivative, and $\alpha_P$ the volumetric thermal expansion coefficient. Equations (1), (2), and (4) constitute the system of governing equations that describe the temporal and spatial evolutions of density $\rho$, velocity $\mathbf{v}$, and temperature $T$ in a near-critical fluid, respectively. To improve the accuracy near the CP, we supplement the above equations with a real-fluid equation of state, $\delta P = (1/\rho\kappa_T)\delta\rho + (\alpha_P/\kappa_T)\delta T$, with the introduction of the isothermal compressibility $\kappa_T$.

All the thermophysical properties involved are evaluated using the NIST reference data with the exception of the bulk viscosity. Given that there is very little experimental data of the bulk viscosity presently available, its determination depends on theoretical predictions [7-10, 12]. Namely, the divergence of the bulk viscosity on the critical isochore in the vanishing-frequency limit is formulated as $\zeta \cong 0.18 \rho c^2 \pi \eta \xi^3 / k_B T$, where $\xi$ is the correlation length [14], and $k_B$ the Boltzmann constant. A finite-difference numerical scheme [15] with a special treatment of the boundary conditions [16] is relied on to solve the equations. The analysis below is based on simulations carried out at various initial distances to the CP (indicated by $\Delta T = T_i - T_c$, with $T_c$=304.13 K) along the critical isochore (that is, $\rho_i = \rho_c$=467.8 kg/m$^3$). The subscript $i$ designates the initial state.

In Fig. 1 we show the numerical results for the continuous heating case of the experiment. Heat of 367 erg was released into the fluid within a period of 200 $\mu$s. Despite the generally fair agreement shown in Fig. 1, a few discrepancies are noticed. First, the predicted wave amplitudes appear to slightly deviate from those observed in the measurement. The theory of sound emission proposed in [12] gives the stepwise density change $\delta\rho/\rho = (1/cT)(\partial T/\partial P)_s Q = (1/\rho c)(\alpha_P/C_P) Q = 2.802 \times 10^{-7}$ for $\Delta$T=30 mK, and 2.591 × $10^{-7}$ for $\Delta T$=150 mK, where $Q$ is the applied heat flux (defined in unit of $W/m^2$). Our simulation seems to better fit with the theoretical estimation than the measurement. Second, a more severe damping is shown in the experimental data for $\Delta T$=30 mK. As the CP is approached, the solid wall can no longer be considered perfectly isothermal due to the increasing fluid effusivity [9], leading to possible discrepancy between the simulation and the experiment.

The fluid's response to the continuous heating is explained as follows. As the heating continues, ultra-thin boundary layers of elevated temperature and pressure grow on the surfaces of the heater, whose strong expansion emits two compressive acoustic waves (in opposite directions). Driving organized flow motions, these thermo-acoustic waves carry the heat to the interior of the fluid. Upon reflection at the rigid isothermal wall, fluid velocity must decelerate to zero, which leads to the formation of a kinetic boundary layer. Similarly, a thermal boundary layer is developed, where temperature varies from the bulk to the fixed value at the boundary. With the vanishingly small thermal diffusivity near the CP, the heat that the wave carries cannot be efficiently diffused through the boundary via thermal conduction. As a result, a large portion of the energy is reflected from the boundary, whose superposition with the incoming wave thus increases the temperature and pressure. Shown in the inset of Fig. 1 is such reflection pattern, where the reflected pressure waves seem to directly "stack" upon the incoming waves.

Even on such a short time scale, the effect of the critically-diverging bulk viscosity can be clearly seen. As shown in the inset of Fig. 1, pronounced pressure gradients emerge for $\Delta T$=30 mK close to the walls. Analogous to Carlès's findings on a much longer time scale [10], the appearance of a pressure gradient is vital to counterbalance the large viscous stresses at the boundary layer. Also noted is that the relatively low pressure inside the boundary layer ensures that no flow reversal from the boundary layer to the bulk is developed, which is in stark contrast to what we see with the pulse heating.

The simulation for the case of pulse heating produces more interesting results. Figure 2 displays the calculated density variations at the cell center, which correspond to the measurement presented in Fig. 4 of [12]. The wave shape varies with respect to the different heating durations. By a closer examination of Fig. 2(b), an intriguing feature arises: stepwise baseline increments appear in the curve for $\Delta T$=60 mK, which are absent in the curve for $\Delta T$=500 mK. Such a near-critical phenomenon has not been captured by previous studies. Note that for the time considered, the pulse has not yet been greatly broadened by the bulk damping. We thus assume the unexpected density increase to be due more to the boundary effect than the bulk attenuation. Detailed analysis of the reflection mechanism demonstrates that both of the finite heat input and the diverging bulk viscosity contribute to this anomaly.

Compared with the continuous heating, the reflection pattern with the pulse heating is found to be quite complex. Since the heating duration [only 10 $\mu$s in Fig. 2(b)] is far less than the typical acoustic time ($t_{ac}\approx$70 $\mu$s), the thermo-acoustic wave emitted at the heater assume the shape of a short pulse; whereas long continuous waveform is generated with the continuous heating (see the inset of Fig. 1). The finite waveform means that, when the pulse passes, the local fluid always experiences strong compression (due to the rising edge of the pulse) and then expansion (due to the falling edge of the pulse) within a very short period of time, and tends to return to the undisturbed state afterwards. In contrast, the continuous heating only creates compressive effect in the fluid. When the pulse arrives at the wall, the fluid similarly engages in two successive reactions—contraction followed by rarefaction. Unable to proceed further in face of a closed end, the flow motion has to reverse its course during the rarefaction part. This is notably different from continuous heating as the uninterrupted expansion of the thermal diffusion layers at the heater forces the fluid to always flow towards the wall. The large viscous effect near the boundary gives rise to excess attenuation. The irreversible loss of momentum and energy causes fluid mass to be “trapped” in the boundary layer.

Next we zoom in on a small portion in the immediate vicinity of the solid wall. Figure 3 illustrates the snapshots of the temperature, pressure, density, and velocity profiles next to the left boundary in the wake of the pulse reflection. Large mass density develops in the boundary layer as predicted for $\Delta T$=60 mK [see Fig. 3(c)] due to the strongly diverging bulk viscosity. Note that mass accumulation also takes place for $\Delta T$=500 mK, but somewhat outside the layer and of a much smaller scale. A temperature “dip” appears where the elevated mass density stands, as shown in Fig. 3(a). From an energy standpoint, the isothermal boundary conditions demands, during rarefaction, the formation of a relatively low-temperature region to draw heat from the ambient, which counteracts the cooling effect caused by the outgoing fluid and preserves local thermal equilibrium. Due to lack of flow motion after the pulse reflection [see Fig. 3(d)], the local temperature inhomogeneity needs a long time to relax by thermal diffusion alone. Interesting though, despite the even smaller thermal diffusivity, the negative “dip” in the curve for $\Delta T$=60 mK is barely noticeable in Fig. 3(a), which suggests a new mechanism might be at work.

Recall that pressure change is closely related to density and temperature variations. As shown in Fig. 3(b), the increased mass density translates into pressure increase at the isothermal boundary. A significant pressure gradient, as a result, appears within the boundary layer for $\Delta T$=60 mK. The relatively high pressure inside the layer induces acoustic expansion of the boundary layer due to the high compressibility near the CP, which is evidenced in the apparent flow motion near the wall [see Fig. 3(d)]. The convection accelerates the thermal relaxation, which explains the almost “dip-free” temperature profile shown in Fig. 3(a). Such a process bears the hallmarks of the PE—thermo-acoustic convection caused by strong expansion of a boundary layer—and should be termed as the *secondary piston effect*. The heat extracted from the ambient is transported to the bulk via this mechanism, which is responsible for the gradual thermalization that we have shown in Fig. 2(b).

In Fig. 4, we plot the time evolution of heat flux at the right boundary, which confirms the duality of the pulse reflection. Heat outflow occurs during the contraction; whereas heat inflow occurs during the expansion. It is noted that the rarefaction part becomes more distinct on approaching the CP. As shown in the inset, we can clearly see the influence of the secondary piston effect. The magnitude of heat flow at the isothermal boundary is rapidly declining because the local thermal imbalance is being relaxed by the fast process.

In summary, we have explored the thermo-acoustic wave emission and reflection pattern on the fast time scale by takings into account the critical divergence of the bulk viscosity, along with other real-fluid properties. The numerical simulation yielded remarkable agreement with the experiment of Miura et al. [12]. New features of the piston effect accrued from our study. The analysis revealed the different responses to the perturbations of continuous heating and pulse heating in a near-critical fluid, and the role of the diverging bulk viscosity in the new process that is identified as the secondary piston effect. There are a few future problems that need our attention. One of them is to employ a more realistic boundary condition that involves solid walls of finite thermal conductivity and thermal capacity as the diverging fluid effusivity could further complicate the boundary layer behavior. Research in this direction might lead to new discoveries.

This research is supported by National Natural Science Foundation of China under contract No. 50776057.

FIG. 1. (Color online) Comparison between the calculated density variation as a function of time at the cell center ($L$=10.3 mm) under the continuous heating in the region $0<t<0.2$ ms and the measurement, red dash dot: $\Delta T$=150 mK; black solid: $\Delta T$=30 mK. The experimental data from Ref. [12] are also included, red cross: $\Delta T$=150 mK; black circle: $\Delta T$=30 mK. As soon as the heater is turned off, the strong expansion of the thermal boundary layers stops almost immediately, which thus gives rise to the trapezoidal shape of the density variation as shown in the figure for $t$>200 $\mu$s. Inset: pressure spatial distributions under the continuous heating prior to and after the thermo-acoustic wave reflection at the right boundary, black solid: $\Delta T$=30 mK at $t$=40 $\mu$s; black dash dot: $\Delta T$=30 mK at $t$=110 $\mu$s; red solid: $\Delta T$=150 mK at $t$=40 $\mu$s; red dash dot: $\Delta T$=150 mK at $t$=110 $\mu$s. The arrows indicate the direction of the thermo-acoustic wave propagation.

FIG. 2. (Color online) Midpoint density change as a function of time under the pulse heating, with parameters specified as (a) $L$=5.5 mm with energy input of 53.6 erg, red dash dot: $\Delta T$=500 mK; black

solid: $\Delta T$=100 mK, and (b) $L$=10.3 mm with energy input of 129 erg, red dash dot: $\Delta T$=500 mK; black solid: $\Delta T$=60 mK. A heating duration of 4.5 $\mu$s is chosen in (a) such that the two compressive waves emitted from the heater can be mutually distinguishable from each other; whereas the single-peak wave profile does not appear until the heating period is adjusted to 10 $\mu$s (as opposed to 7 $\mu$s stated in [12]). Note that the thermo-acoustic waves assume finite waveforms, which is in accordance with the pulse heat input.

FIG. 3. (Color online) Plots of the spatial profiles for (a) temperature, (b) pressure, (c) density, and (d) velocity close to the left boundary under the pulse heating, black solid: $\Delta T$=60 mK at $t$=440 $\mu$s; red dash dot: $\Delta T$=500 mK at $t$=400 $\mu$s. The curves are plotted for different time instants with respect to the initial temperatures so as to compensate for the sound velocity differences.

FIG. 4. (Color online) Heat flux history at the right boundary layer under the pulse heating, black solid: $\Delta T$=60 mK; red dash dot: $\Delta T$=500 mK. Inset: enlarged bottom portion between wave arrivals. Positivity indicates heat outflow from the fluid to the ambient.

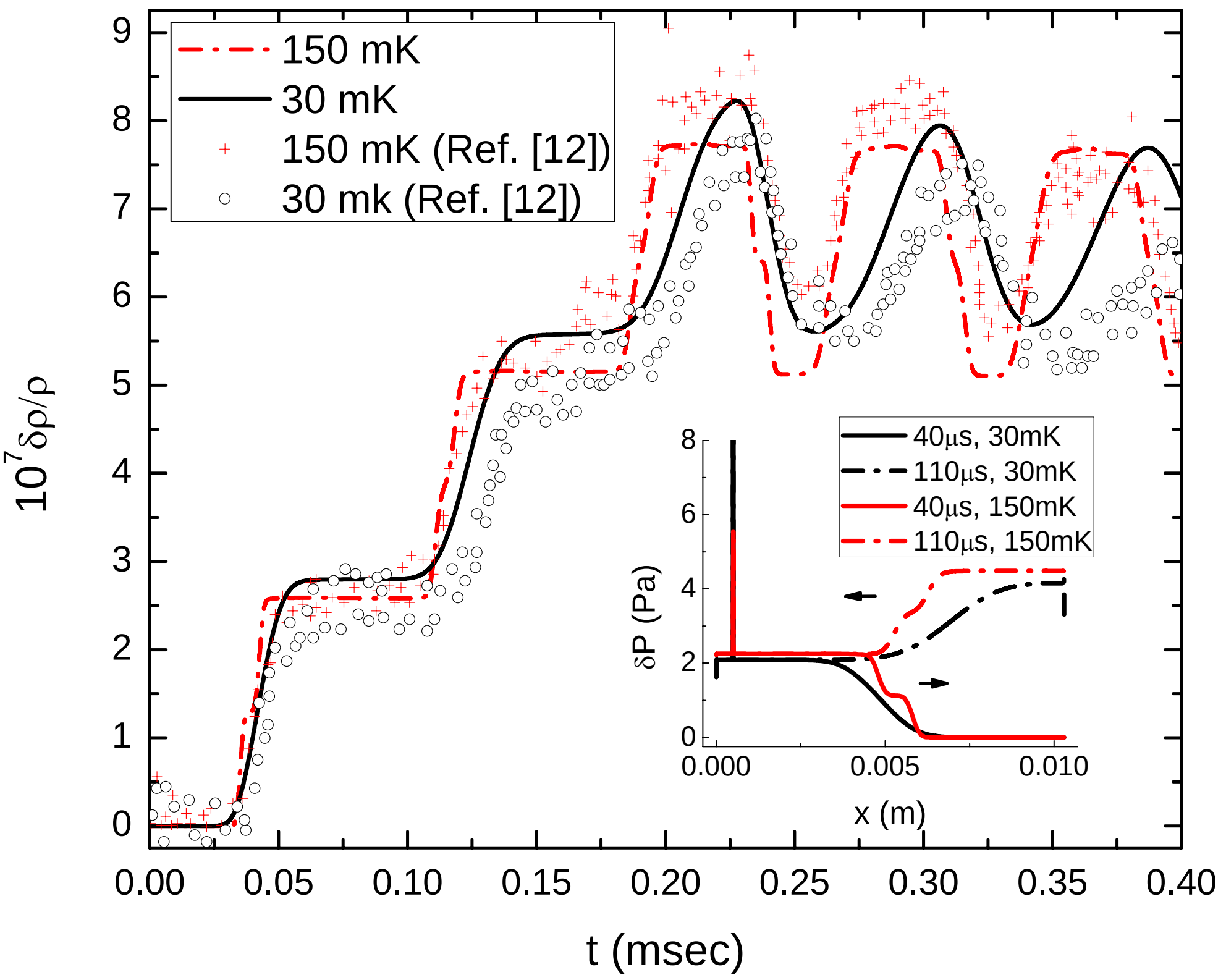

FIG. 1.

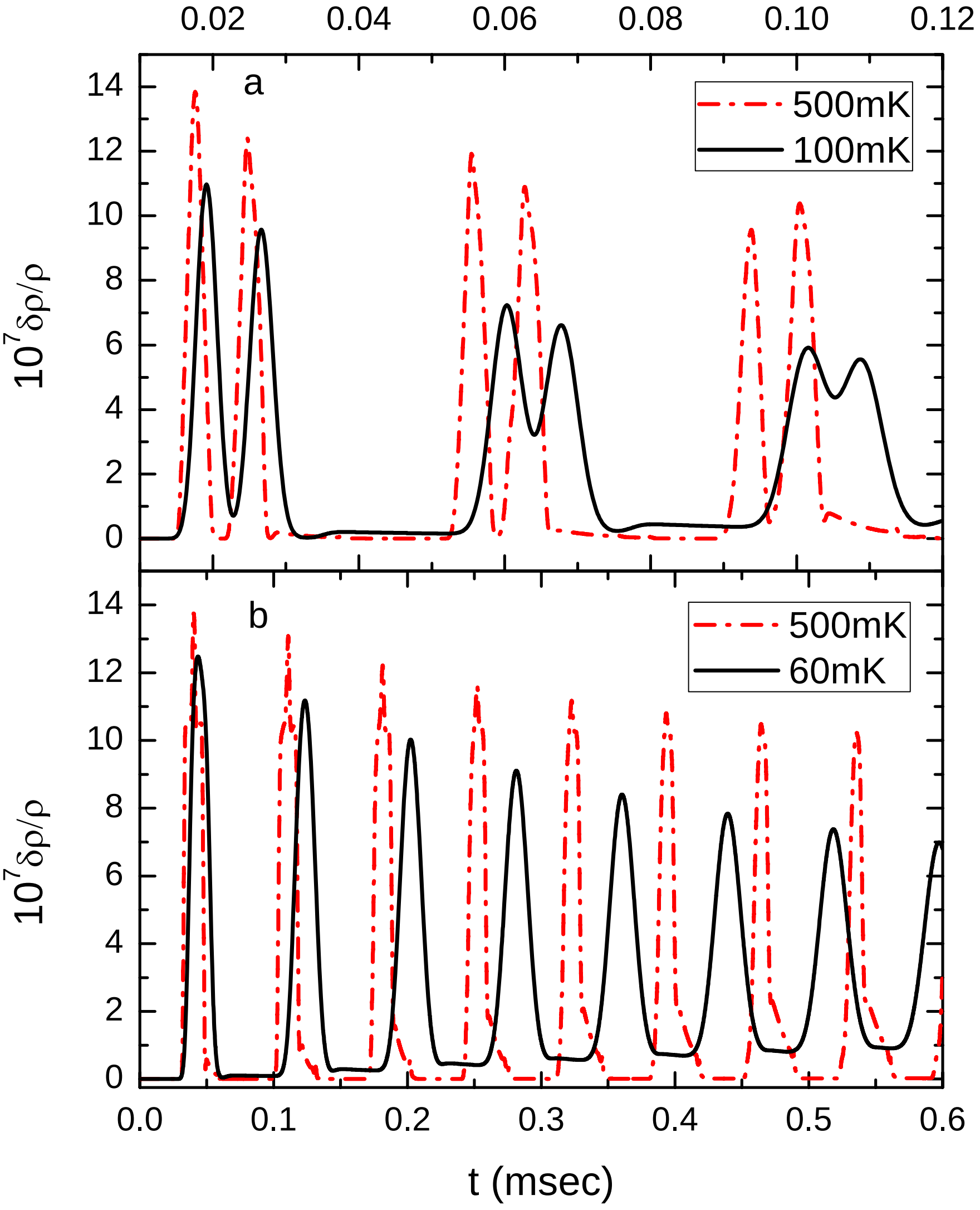

FIG. 2.

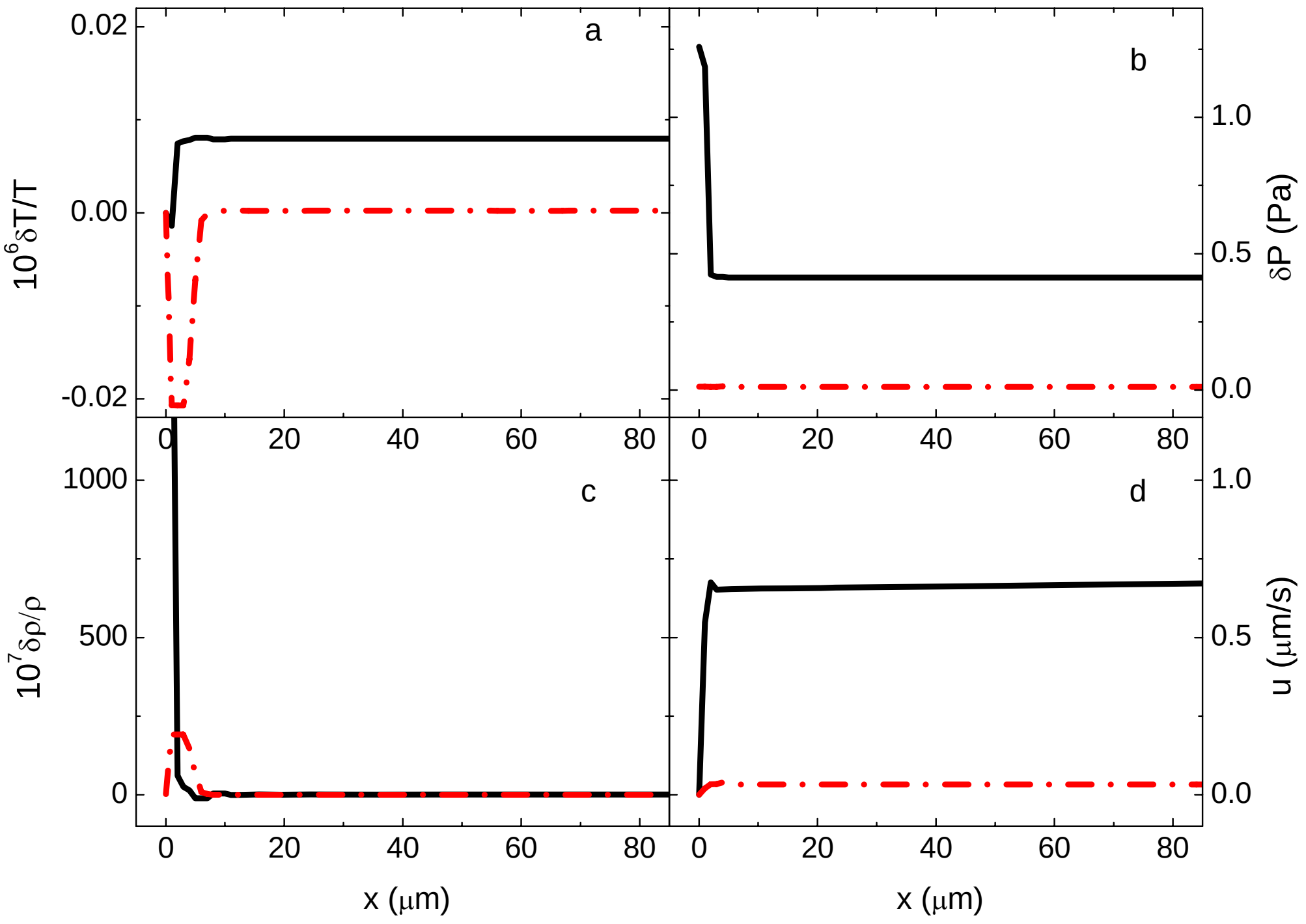

FIG. 3.

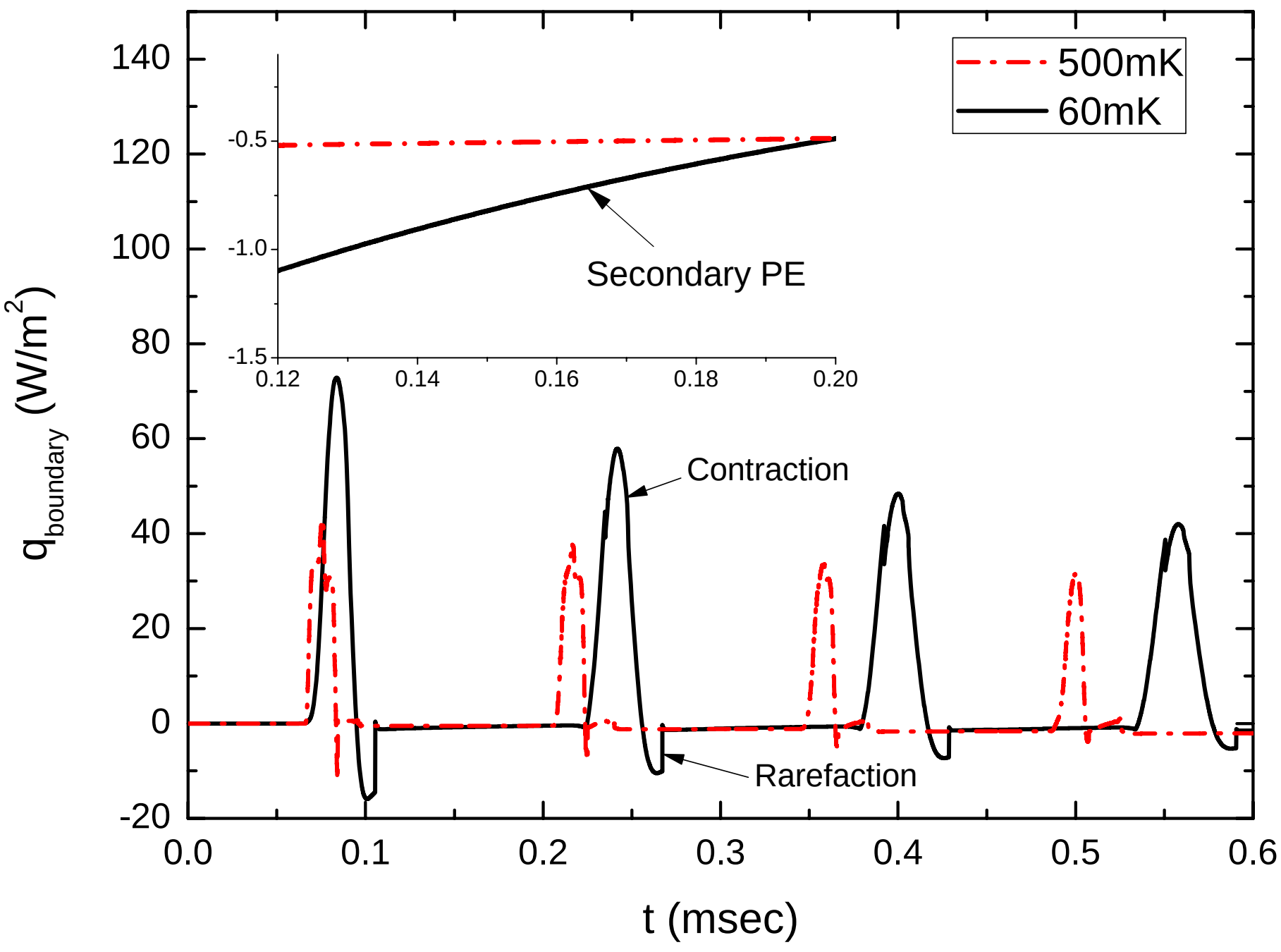

FIG. 4.